\documentclass[12pt]{article}
\newcommand{\sect}[1]{\setcounter{equation}{0}\section{#1}\indent}

\def\[{{[}}

\def\CC{{\cal C}}
\def\CP{{\cal P}}

\def\nn{\nonumber}
\def\1{1\hspace{-1.5mm}1}
\newcommand{\bea}{\begin{eqnarray}}
\newcommand{\eea}{\end{eqnarray}}
\newcommand{\bref}[1]{(\ref{#1})}
\renewcommand{\nn}{\nonumber}
\newcommand{\hepth}[1]{{\sf hep-th/{#1}}}

\begin{document}

\thispagestyle{empty}
\hfill March 29, 2002\par
\hfill KEK-TH-814\par
\hfill TOHO-CP-0271\par
\hfill hep-th/0204002
\vskip 20mm
\begin{center}
{\Large\bf Super-PP-wave Algebra\\
from\\
\vspace{3mm}
Super-AdS$\times S$
Algebras in Eleven-dimensions}
\vskip 6mm
\medskip

\vskip 10mm
{\large Machiko\ Hatsuda,~Kiyoshi\ Kamimura$^\dagger$~and~Makoto\ Sakaguchi 
}\par
\medskip
{\it 
Theory Division,\ High Energy Accelerator Research Organization (KEK),\\
\ Tsukuba,\ Ibaraki,\ 305-0801, Japan} \\
{\it 
$~^\dagger$ 
 Department of Physics, Toho University, Funabashi, 274-8510, Japan}\\
\medskip\medskip
{\small\sf E-mails:\ mhatsuda@post.kek.jp, kamimura@ph.sci.toho-u.ac.jp,
Makoto.Sakaguchi@kek.jp} 
\medskip
\end{center}

\vspace{15mm}

\begin{abstract}
Maximally supersymmetric spacetime algebras in eleven-dimensions,
which are the isometry superalgebras of
Minkowski space,
AdS$_7\times S^4$,
AdS$_4\times S^7$
and pp-wave background,
are related by In\"on\"u-Wigner contractions.
The super-AdS$_{4(7)}\times S^{7(4)}$ algebras
allow to introduce two contraction parameters,
the one for the flat limit to the super-Poincar\'e algebra
and the other for a Penrose limit to the super-pp-wave algebra. 
Under these contractions supersymmetries are maintained
because the Jacobi identity of three supercharges holds
for any values of contraction parameters.
\end{abstract}
\vspace{10mm} 
\par\par
\newpage

\setcounter{page}{1}
\parskip=7pt

\section{Introduction}

Recently, maximally supersymmetric pp-wave (plane-fronted gravitational
wave with parallel rays)
backgrounds of supergravity theories in eleven- and ten-dimensions
have attracted great interests.
In \cite{KG}, maximally supersymmetric backgrounds in eleven-dimensions 
were classified into four types;  flat Minkowski space (and their toroidal 
compactifications), AdS$_4\times S^7$, AdS$_7\times S^4$ and
the super-pp-wave background.
This super-pp-wave background in eleven-dimensions
was reconsidered recently in \cite{FP}.
For the type-IIB supergravity,
a super-pp-wave background was discovered
in \cite{BFHP;IIB}.\footnote{
For other dimensions see \cite{KG;positive} and \cite{Me}.
}
The type-IIB superstrings in this background were shown to be exactly solvable
\cite{M} in spite of the presence of the RR five-form field strength.
This model is expected to provide some hints for
the study of superstrings on more general backgrounds
\cite{MT}\cite{BP}\cite{Ber}.

The Penrose limit plays a central role in these recent studies of the pp-wave 
backgrounds.
It was known that any solution of Einstein gravity admits plane-wave backgrounds
in the Penrose limit \cite{P}.
This was extended to the solutions of supergravities in \cite{G}
and was studied in the context of the Wess-Zumino-Witten models in \cite{WZW}.
It was shown that the super-pp-wave background can be derived by the Penrose limit 
from the super-AdS$\times S$ backgrounds in \cite{BFHP;Penrose} \cite{BFP}
and this was investigated further in \cite{Penrose limit}.
In addition, the full supersymmetry
algebra of the super-pp-wave background of the type-IIB theory
was shown to be derived by an In\"on\"u-Wigner (IW) contraction \cite{IW}
from the super-AdS$_5\times S^5$ algebra of SU(2,2$|$4) in \cite{HKS}.
The Penrose limit was recognized to be important to explore
the AdS/CFT correspondence \cite{AdS/CFT} beyond massless string modes in 
\cite{BMN}.
This line was pursued further in
\cite{YM}.

In this paper, we consider the Penrose limit from an algebraic point of view,
which is coordinate-independent and manifestly supersymmetric.
We show that the isometry superalgebra of the super-pp-wave background
in eleven-dimensions
is obtained from the super-AdS$_{4(7)}\times S^{7(4)}$ algebras
by IW contractions.
The supersymmetries are maintained during contractions
because the Jacobi identity of three supercharges holds
for any values of contraction parameters.
In section 2, the pp-wave algebra is derived from
the AdS$_{4(7)}\times S^{7(4)}$ algebras.
This is extended to the maximally supersymmetric case in section 3.
The last section is devoted to a summary and discussions.

\section{PP-wave Algebra from AdS$_{4(7)}\times S^{7(4)}$ Algebras}

The AdS$_{p+2}\times S^{9-p}$ algebra,
which is the isometry algebra of the AdS$_{p+2}\times S^{9-p}$ space,
is given, in terms of dimensionless 
momenta $P$'s and rotations $J$'s, as
\begin{eqnarray}
\[P_{a},P_{b}]=4\epsilon^2J_{ab},&&
\[P_{a'},P_{b'}]=-\epsilon^2J_{a'b'},\nn\\
\[J_{ab},P_{c}]=2\eta_{bc}P_a,&&
\[J_{a'b'},P_{c'}]=2\eta_{b'c'}P_{a'},\label{AdS algebra;bosonic}\\
\[J_{ab},J_{cd}]=4\eta_{ad}J_{bc},&&
\[J_{a'b'},J_{c'd'}]=4\eta_{a'd'}J_{b'c'},\nn
\end{eqnarray}
where 
$\epsilon^2=1$ for $p=2$ and $\epsilon^2=-1$ for $p=5$.
It is understood that the obvious anti-symmetries of indices
in the left hand side of equalities
in \bref{AdS algebra;bosonic}
are to be implemented on the right hand side
with unit weight.
For $p=2$, this algebra is the AdS$_4\times S^7$ algebra
with the vector index of  AdS$_4$, $a=0,1,2,3$, and that of
S$^{7}$,  $a'=4,...,9,\natural$.
On the other hand, for $p=5$, this algebra is the AdS$_7\times S^4$ algebra
with the vector index of $S^4$,  $a=\natural,1,2,3$,
and that of AdS$_{7}$, $a'=4,...,9,0$.
We use the metric $\eta_{\mu\nu}$ with $\eta_{00}=-1$ otherwise $+1$.

The symmetry group is SO$(p+1,2)\times$SO$(10-p)$
which has a flat limit to ISO$(p+1,1)\times$ISO$(9-p)$
by the following IW contraction
\begin{enumerate}
\item Rescale the translation generators $P$'s as
\bea
P_a\to RP_a,~~~~~P_{a'}\to R' P_{a'}.
\label{scalingR}
\eea
where $R/2$ and $R'$ are the radii of the AdS$_4$($S^4$)
and that of $S^{7}$(AdS$_7$), respectively.

\item Then take $R\to \infty$ and $R'\to \infty$.  
\end{enumerate}  
In this limit, $P$'s become the linear momenta and $J$'s are the Lorentz
generators.
For the maximal spacetime supersymmetry, $R$ and $R'$ must satisfy
\begin{eqnarray}
R=R',~~~
R^2=2\rho R_S^2=\frac{2}{\rho}R_{AdS}^2,~~~
\rho=\left\{
  \begin{array}{ll}
    1/2   & \mbox{for}~~ p=2    \\
     2  & \mbox{for}~~ p=5    \\
  \end{array}
\right.
\label{R}
\end{eqnarray}
where $R_S$ and $R_{AdS}$ are the radii of $S^{9-p}$ and that of AdS$_{p+2}$,
respectively. 
\bigskip

Besides the flat limit, as any other metrics,
the AdS$\times S$ metric allows the Penrose limit \cite{P}
giving a plane wave metric. 
The Penrose limit 
can be understood as an IW contraction of the AdS$_{p+2}\times S^{9-p}$ algebra
into the pp-wave algebra. 
\begin{enumerate}
\item Define the light cone components of the momenta $P$'s 
and {\it boost} generators $P^*$'s as
\bea
&&P_{\pm}\equiv\frac{1}{\sqrt{2}}( P_\natural\pm P_0),~~~
P_{m}~=~(P_i,P_{i'}),\nn\\
&&P_m^*=
\left\{
  \begin{array}{ll}
    (P_i^\ast=J_{i0}, P_{i'}^\ast=J_{i'\natural})   &\mbox{ for }p=2   \\
    (P_i^\ast=J_{i\natural}, P_{i'}^\ast=J_{i'0})     & \mbox{ for }p=5    \\
  \end{array}
\right.
\label{lc}
\eea
where $i=1,2,3,~~i'=4,...,9,$ and $m=(i,i').$
\item
Suppose the plane-wave propagates along $x_+$ {\it time} direction.
The transverse translation and { boost} generators
are rescaled with 
a dimensionless parameter $\Omega$ 
as
\bea
P_+\to \frac{1}{\Omega^2}P_+~~,
P_m\to \frac{1}{\Omega}P_m~~,~~
P_m^*\to \frac{1}{\Omega}P_m^*.
\label{penomg}
\eea
\item Then take $\Omega\to 0$ limit.  
\end{enumerate}

To see them explicitly the AdS$_{p+2}\times S^{9-p}$
algebra (\ref{AdS algebra;bosonic}) 
is rescaled,  following to \bref{scalingR} and \bref{penomg}, as
\bea
\[P_i,P_+]=2\sqrt{2}\frac{\epsilon^2\Omega^2}{R^2}P_i^\ast,&&
\[P_{i'},P_+]=-\frac{\epsilon^2\Omega^2}{\sqrt{2}{R}^2}P_i^\ast,\nn\\
\[P_{i}^\ast, P_+]=-\frac{1}{\sqrt{2}}\epsilon^2\Omega^2P_i,&&
\[P_{i'}^\ast, P_+]=\frac{1}{\sqrt{2}}\epsilon^2\Omega^2P_{i'},\nn\\
\[P_{i},P_-]=-2\sqrt{2}\frac{1}{R^2}P_{i}^\ast,&&
\[P_{i'},P_-]=-\frac{1}{\sqrt{2}{R}^2}P_{i'}^\ast,\nn\\
\[P_{i}^\ast,P_-]=\frac{1}{\sqrt{2}}P_i,&&
\[P_{i'}^\ast,P_-]=\frac{1}{\sqrt{2}}P_{i'},\nn\\
\[P_{i}^\ast,P_{j}]=\frac{1}{\sqrt{2}}\eta_{ij}(\epsilon^2\Omega^2P_--P_+),&&
\[P_{i'}^\ast,P_{j'}]=-\frac{1}{\sqrt{2}}\eta_{i'j'}(\epsilon^2\Omega^2P_-+P_+),
\nn\\
\[P_i,P_j]=4\frac{\epsilon^2\Omega^2}{R^2}J_{ij},&&
\[P_{i'},P_{j'}]=-\frac{\epsilon^2\Omega^2}{{R}^2}J_{i'j'},\nn\\
\[P_{i}^\ast, P_{j}^\ast]=\epsilon^2\Omega^2J_{ij},&&
\[P_{i'}^\ast, P_{j'}^\ast]=-\epsilon^2\Omega^2J_{i'j'},\nn\\
\[J_{ij},P_{k}]=2\eta_{jk}P_i,&&
\[J_{i'j'},P_{k'}]=2\eta_{j'k'}P_{i'},\nn\\
\[J_{ij},P_{k}^\ast]=2\eta_{jk}P_i^\ast,&&
\[J_{i'j'},P_{k'}^\ast]=2\eta_{j'k'}P_{i'}^\ast,\nn\\
\[J_{ij},J_{kl}]=4\eta_{il}J_{jk},&&
\[J_{i'j'},J_{k'l'}]=4\eta_{i'l'}J_{j'k'},
\label{bosonRomg}
\eea
where we used relation $R=R'$ of (\ref{R}) for simplicity.
By taking $\Omega\to 0$ limit,
the $\epsilon$ dependence, which distinguishes the $p=2$ case from the  $p=5$ case,
disappears and
\bref{bosonRomg} becomes the unique plane wave algebra in eleven-dimensions, 
\bea
\[P_{i},P_-]=-2\sqrt{2}\frac{1}{R^2}P_{i}^\ast,&&
\[P_{i'},P_-]=-\frac{1}{\sqrt{2}{R}^2}P_{i'}^\ast,\nn\\
\[P_m^\ast,P_-]=\frac{1}{\sqrt{2}}P_m,&&
\[P_m^\ast,P_n]=-\frac{1}{\sqrt{2}}\eta_{mn}P_+,\nn\\
\[J_{mn},P_{p}]=2\eta_{np}P_m,&&
\[J_{mn},P_{p}^\ast]=2\eta_{np}P_m^\ast,\nn\\
\[J_{mn},J_{pq}]=4\eta_{mq}J_{np}.&&
\label{ppalge}
\eea
This is the symmetry algebra of the pp-wave metric \cite{KSHM}. 
The Poincar\'e algebra of ISO($10,1$) group
is obtained by taking flat limit $R\to 0$ in (\ref{ppalge})
and by making manifest restored symmetry generators such as
$J_{ij'}$, $J_{\pm i}$, etc.

\sect{
Super-PP-wave algebra from
Super-AdS$_{4(7)}\times S^{7(4)}$ Algebras
}

We extend the previous analysis
to the supersymmetric case 
and show the maximally supersymmetric eleven-dimensional
pp-wave algebra \cite{FP}
is obtained by IW contractions of the super-AdS$_{p+2}\times S^{9-p}$ algebras.
In this case
the Jacobi identities of the superalgebra give a restriction 
on the radii of AdS and $S$, $R_{AdS}$ and $R_{S}$,
as $R_{AdS}=\frac{1}{2}R_{S}$ for $p=2$ and 
$R_{AdS}={2}R_{S}$ for $p=5$, as in (\ref{R}).
In eleven-dimensions, the supersymmetry generators $Q_\alpha$
are $32$ Majorana spinors.
The gamma matrices $\Gamma^{\mu},~(\mu=0,1,..,9,\natural)$, 
in eleven-dimensions are composed in terms of
those of AdS$_{4}$, $\gamma^a,~({a}=0,1,2,3)$, and of $S^{7}$, $\gamma^{a'},~
({a'}=4,5,..,9,\natural)$,
or in terms of those of $S^{4}$, $\gamma^a,~({a}=\natural,1,2,3)$,
and of AdS$_{7}$, 
$\gamma^{a'},~({a'}=4,5,..,9,0)$,
as
\footnote{
For the gamma-matrices in four-dimensions,
we use
a Majorana representation for AdS$_4$ ($a=0,1,2,3$)
and a pseudo-symplectic representation for $S^4$ ($a=\natural,1,2,3$)
which satisfy, in both cases,
\begin{eqnarray*}
\tilde C=-C,~~~
\tilde \gamma^a=-C\gamma^aC^{-1},~~~
\{\gamma^{a},\gamma^{b}\}=2\eta^{ab}
\end{eqnarray*}
and $C\gamma^{a_1\cdots a_N}$ is symmetric iff $N=1,2~mod~4 $.
For the gamma-matrices in  seven-dimensions,
we use a pseudo-Majorana representation for $S^7$ ($a'=4,5,...,9,\natural$)
and a symplectic Majorana representation for AdS$_7$
($a'=4,5,...,9,0$) which satisfy, in both cases,
\begin{eqnarray*}
\tilde C'=+C',~~~
\tilde \gamma^{a'}=-C'\gamma^{a'}C'^{-1},~~~
\{\gamma^{a'},\gamma^{b'}\}=2\eta^{a'b'}
\end{eqnarray*}
and $C'\gamma^{a_1'\cdots a_N'}$ is symmetric iff $N=0,3~mod~4 $.
}
\bea
&&\Gamma^a=\gamma^a\otimes {\1},~~~
\Gamma^{a'}=\gamma_5\otimes \gamma^{a'},~~~~
\gamma_5=\left\{
  \begin{array}{ll}
   i\gamma^{0123} &\mbox{for }~ p=2   \\
   \gamma^{\natural 123} & \mbox{for }~p=5   \\
  \end{array}
\right.
.
\eea
They satisfy $\{\Gamma^\mu,\Gamma^\nu\}=2\eta^{\mu\nu}$.
The charge conjugation matrix ${\cal C}$ in eleven-dimensions is taken as
\bea
{\cal C}=C\otimes C'
\eea
where $C$ and $C'$ are 
the charge conjugation matrices for AdS$_4$ ($S^4$)
and $S^7$ (AdS$_7$) spinors respectively.

\bigskip

The bosonic part of the super-AdS$_{p+2}\times S^{9-p}$ algebra
is \bref{AdS algebra;bosonic}.
In addition to it, the odd generators $Q_\alpha$ satisfy
\bea
\[P_a,Q]=i\epsilon Q\gamma_5\gamma_a\otimes \1,&&
\[P_{a'},Q]=\frac{i}{2}\epsilon Q\1\otimes \gamma_{a'},\nn\\
\[J_{ab},Q]=\frac{1}{2}Q\gamma_{ab}\otimes\1,&&
\[J_{a'b'},Q]=\frac{1}{2}Q\1\otimes\gamma_{a'b'},\nn
\eea
\bea
\{Q,Q\}&=&-2C\gamma^a\otimes C'P_a
 -2C\gamma_5\otimes C'\gamma^{a'}P_{a'}
\nn\\&&
 -2i\epsilon C\gamma_5\gamma^{ab}\otimes C' J_{ab}
 +i\epsilon C\gamma_5\otimes C'\gamma^{a'b'} J_{a'b'}.
\label{QQ}
\eea
The Jacobi identities are satisfied as long as $\epsilon^2=1$ for $p=2$
and $\epsilon^2=-1$ for $p=5$.
Hereafter, we choose $\epsilon=1$ for $p=2$ and $\epsilon=-i$
for $p=5$, for presentation.

For the present purpose, we rewrite this superalgebra 
in terms of eleven-dimensional covariant gamma matrices, 
$\Gamma$'s, instead of $\gamma$'s.
It is convenient to introduce following matrix  
\bea
I=\Gamma^{123},~~~
\Gamma^{\sharp 123}=\Gamma^\sharp I =-i\epsilon\gamma_5\otimes \1,~~~
\Gamma^\sharp=\left\{
  \begin{array}{ll}
    \Gamma^0   &\mbox{for }~ p=2   \\
    -\Gamma^\natural   &\mbox{for }~ p=5    \\
  \end{array}
\right.
.
\label{gammabar}
\eea

The commutation relations \bref{QQ} for the super-AdS$_{p+2}\times S^{9-p}$
algebra are rewritten as 
\bea
&&\[P_a,Q]=-Q\Gamma^\sharp I\Gamma_a,~~~
\[P_{a'},Q]=-\frac{1}{2}Q\Gamma^\sharp I\Gamma_{a'},~~~
\[J_{mn},Q]=\frac{1}{2}Q\Gamma_{mn}\nn\\
&&\{Q,Q\}=-2\CC\Gamma^\mu P_\mu
 +2\CC\Gamma^\sharp I\Gamma^{ab}J_{ab}
 -\CC\Gamma^\sharp I\Gamma^{a'b'}J_{a'b'}.
\label{QQG}
\eea

Associating with the rescaling of $P$'s by $R$ as in \bref{scalingR} with (\ref{R}),
the dimensionless supercharges $Q_\alpha$ are rescaled as
\bea
P_a\to RP_a~~,~~P_{a'}\to R P_{a'}
\label{scalingR2}~~,~~
Q_\alpha\to \sqrt{R}~Q_\alpha.
\label{scalingQR}
\eea
Corresponding to the rescaling of bosonic generators \bref{penomg},
the components of the supercharges must be rescaled with proper weights.
To do this,
we decompose supercharges $Q$
as
\bea
Q&=&Q_{+}+Q_{-},~~~~Q_\pm=Q_\pm\CP_\pm,
\eea
using
the light cone projection operators
\begin{eqnarray}
\CP_\pm=\frac{1}{2}\Gamma_\pm\Gamma_\mp,~~~
\Gamma_\pm\equiv\frac{1}{\sqrt{2}}(\Gamma_\natural\pm\Gamma_0).
\end{eqnarray}

In order to obtain the well defined limit of 
the super-AdS$_{p+2}\times S^{9-p}$ algebra,
the supercharges turn out to be rescaled as
\bea
Q_+\to \frac{1}{\Omega}~Q_+~~,~~Q_-\to Q_-.
\label{penQomg}
\eea

The super-AdS$_{p+2}\times S^{9-p}$
algebra \bref{QQG},
after the rescaling \bref{scalingQR}
by $R$  and \bref{penomg} and \bref{penQomg} by $\Omega$,
becomes
\bea
\[P_+,Q_+]=\frac{\Omega^2\epsilon^2}{2\sqrt{2}R}Q_+I,&&
\[P_-,Q_+]=-\frac{3}{2\sqrt{2}R}Q_+I,\nn\\
\[P_i,Q_-]=\frac{1}{\sqrt{2}R} Q_+\Gamma^-I\Gamma_i,&&
\[P_{i'},Q_-]=\frac{1}{2\sqrt{2}R} Q_+\Gamma^-I\Gamma_{i'},\nn\\
\[P_m^\ast,Q_-]=\frac{1}{2\sqrt{2}}Q_+\Gamma_m\Gamma^-,&&
\[J_{mn},Q_\pm]=\frac{1}{2}Q_\pm\Gamma_{mn}\nn\\
\[P_+,Q_-]=\frac{3\Omega^2\epsilon^2}{2\sqrt{2}R}Q_-I,&&
\[P_-,Q_-]=-\frac{1}{2\sqrt{2}R}Q_-I,\nn\\
\[P_i,Q_+]=-\frac{\Omega^2\epsilon^2}{\sqrt{2}R}Q_-\Gamma^+I\Gamma_i,&&
\[P_{i'},Q_+]=-\frac{\Omega^2\epsilon^2}{2\sqrt{2}R}Q_-\Gamma^+I\Gamma_{i'},
\nn\\
\[P_i^\ast,Q_+]=-\frac{\Omega^2\epsilon^2}{2\sqrt{2}}Q_-\Gamma_i\Gamma^+,&&
\[P_{i'}^\ast,Q_+]=\frac{\Omega^2\epsilon^2}{2\sqrt{2}}Q_-\Gamma_{i'}\Gamma^+,
\nn
\eea
\bea
\{Q_+,Q_+\}&=&
 -2\CC\Gamma^+P_+
 +\frac{\sqrt{2}\Omega^2\epsilon^2}{R}\CC\Gamma^+I\Gamma^{ij}J_{ij}
 -\frac{\Omega^2\epsilon^2}{\sqrt{2}R}\CC\Gamma^+I\Gamma^{i'j'}J_{i'j'},\nn\\
\{Q_-,Q_-\}&=&
  -2\CC\Gamma^-P_-
 -\frac{\sqrt{2}}{R}\CC\Gamma^-I\Gamma^{ij}J_{ij}
 +\frac{1}{\sqrt{2}R}\CC\Gamma^-I\Gamma^{i'j'}J_{i'j'},\nn\\
\{Q_+,Q_-\}&=&
 \left(
 -2\CC\Gamma^mP_m
 -\frac{4}{R}\CC I\Gamma^iP_i^\ast
 -\frac{2}{R}\CC I\Gamma^{i'}P_{i'}^\ast
 \right)\CP_- .
\label{susyomeQQ}
\eea
It is important that negative power terms of $\Omega$
disappear owing to the presence of the light cone projections.
Therefore we can take the consistent Penrose limit $\Omega\to 0$
of the algebra to obtain
\bea
\[P_-,Q_+]=-\frac{3}{2\sqrt{2}R}Q_+I,&&
\[P_-,Q_-]=-\frac{1}{2\sqrt{2}R}Q_-I,
\nn\\
\[P_i,Q_-]=\frac{1}{\sqrt{2}R} Q_+\Gamma^-I\Gamma_i,&&
\[P_{i'},Q_-]=\frac{1}{2\sqrt{2}R} Q_+\Gamma^-I\Gamma_{i'},\nn\\
\[P_m^\ast,Q_-]=\frac{1}{2\sqrt{2}}Q_+\Gamma_m\Gamma^-,&&
\[J_{mn},Q_\pm]=\frac{1}{2}Q_\pm\Gamma_{mn},\nn
\eea
\bea
\{Q_+,Q_+\}&=&
 -2\CC\Gamma^+P_+
\nn\\
\{Q_-,Q_-\}&=&
  -2\CC\Gamma^-P_-
 -\frac{\sqrt{2}}{R}\CC\Gamma^-I\Gamma^{ij}J_{ij}
 +\frac{1}{\sqrt{2}R}\CC\Gamma^-I\Gamma^{i'j'}J_{i'j'},\nn\\
\{Q_+,Q_-\}&=&
 \left(
 -2\CC\Gamma^mP_m
 -\frac{4}{R}\CC I\Gamma^iP_i^\ast
 -\frac{2}{R}\CC I\Gamma^{i'}P_{i'}^\ast
 \right)\CP_-
  .
\label{QQpp}
\eea
After the Penrose limit, all $\epsilon$ dependence have disappeared.
This reflects the fact that the super-AdS$_4\times S^7$ background and the 
super-AdS$_7\times S^4$ background reduce into the unique super-pp-wave background
in the Penrose limit.
In this way, we have obtained the super-pp-wave algebra from
the super-AdS$_4\times S^7$ algebra and the super-AdS$_7\times S^4$ algebra.

Furthermore the flat limit to super-Poincar\'e algebra can be taken by
$R\to \infty$ in \bref{ppalge}
and \bref{QQpp}, and by making manifest restored (bosonic) symmetry generators as
was mentioned in the last section.

We have established relations of the maximally supersymmetric
spacetime algebras
in eleven-dimensions; super-Poincar\'e, super-AdS$_4\times S^7$, 
super-AdS$_7\times S^4$
and super-pp-wave, by IW contractions of superalgebras.


\section{Summary and Discussions}

We derived the super-pp-wave algebra form the super-AdS$_{4(7)}\times S^{7(4)}$
algebras in eleven-dimensions by 
IW contractions which correspond to the Penrose limits of 
the super-AdS$_{4(7)}\times S^{7(4)}$ backgrounds.
The differences between the super-AdS$_{4}\times S^{7}$ algebra and
the super-AdS$_{7}\times S^{4}$ algebra,
which are essentially caused from interchanging $0$-component and 
$\natural$-component, 
are shown to disappear after the contraction
resulting to the unique super-pp-wave algebra in eleven-dimensions.
A solution of plane wave is a function of $x_+=t+x_\natural$ and
it is invariant under interchanging $t$ and $x_\natural$.
The Penrose limit brings to such a plane wave space.  
This naturally explains why the maximally supersymmetric pp-wave algebra
is unique
in spite of the presence of two distinct super-AdS algebras in eleven-dimensions.
This property will be deeply related to the T-duality discussed in \cite{G}. 
Supersymmetries are maintained during contractions because the Jacobi identities
hold for any values of contraction parameters, $R$ and $\Omega$,
and even for their limits $R\to \infty$ and $\Omega\to 0$.

It is noted that one can relate
 the super-pp-algebra,
\bref{ppalge} and \bref{QQpp},
to the one
obtained in \cite{FP}.
To do this, we rewrite generators as
\begin{eqnarray}
&&P_\pm\to3\sqrt{2}e_\pm,~~~
P_m\to3\sqrt{2}e_m,
\nn\\&&
P_i^\ast\to\frac{9\sqrt{2}}{2\mu^2}e_i^\ast,~~~
P_{i'}^\ast\to\frac{18\sqrt{2}}{\mu^2}e_{i'}^\ast,~~~
Q_\pm\to\sqrt{6\sqrt{2}}Q_\pm
\end{eqnarray}
where $\mu=1/R$.
Under this, the super-pp-wave algebra
(\ref{ppalge}) and (\ref{QQpp})
turns out to be
\begin{eqnarray}
&&\[e_m,e_-]=-e_m^\ast,~~~
\[e_{i}^\ast,e_-]=\frac{\mu^2}{9}e_{i},~~~
\[e_{i'}^\ast,e_-]=\frac{\mu^2}{36}e_{i'},\nn\\
&&\[e_i^\ast,e_j]=-\frac{\mu^2}{9}\eta_{ij}e_+,~~~
\[e_{i'}^\ast,e_{j'}]=-\frac{\mu^2}{36}\eta_{i'j'}e_+,\nn\\
&&\[J_{mn},e_p]=2\eta_{np}e_m,~~~
\[J_{mn},e_p^\ast]=2\eta_{np}e_m^\ast,~~~
\[J_{mn},J_{pq}]=4\eta_{np}J_{mq},\nn\\
&&\[e_-,Q_+]=-\frac{\mu}{4}Q_+I,,~~~
\[e_-,Q_-]=-\frac{\mu}{12}Q_-I,~~~
\[e_i,Q_-]=\frac{\mu}{6}Q_+\Gamma^-I\Gamma_i,\nn\\
&&
\[e_{i'},Q_-]=\frac{\mu}{12}Q_+\Gamma^-I\Gamma_{i'},~~~
\[e_i^\ast,Q_-]=\frac{\mu^2}{18}Q_+\Gamma_i\Gamma^-,~~~
\[e_{i'}^\ast,Q_-]=\frac{\mu^2}{72}Q_+\Gamma_{i'}\Gamma^-,\nn\\
&&\{Q_+,Q_+\}=
 -\CC\Gamma^+e_+,\nn\\
&&\{Q_-,Q_-\}=
 -\CC\Gamma^-e_-
 -\frac{\mu}{6}\CC\Gamma^-I\Gamma^{ij}J_{ij}
 +\frac{\mu}{12}\CC\Gamma^-I\Gamma^{i'j'}J_{i'j'},\nn\\
&&\{Q_+,Q_-\}=
 -\CC\Gamma^me_m
 -\frac{3}{\mu}\CC I\Gamma^ie_i^\ast
 -\frac{6}{\mu}\CC I\Gamma^{i'}e_{i'}^\ast.
 \label{FP}
\end{eqnarray}
and is the superalgebra in \cite{FP}
\footnote{  
In the algebra given in \cite{FP},
$Q_+$ and $Q_-$ 
are exchanged
in the commutation relations of
$\[e^\ast_m,Q_-]$ and $\[e_m, Q_-]$.
It apparently breaks the $\Omega$-grading property
which played the central role in the Penrose limit as explained in this paper. }.
Our presentation reveals the way to take the flat limit while
the flat limit $\mu \to 0$ is not obvious in the form of (\ref{FP}).

The relation between the super-pp wave algebra
and the AdS$_{4(7)}\times S^{7(4)}$ algebra is useful
for constructing mechanical actions 
of M-branes in the super-pp wave background.
Any form field in the former can be derived from the corresponding form in
the latter.
This approach will make the (super)symmetry
of the pp-wave systems manifest.


\end{document}